\documentclass[runningheads]{llncs}
\usepackage{graphicx}
\usepackage{caption}
\usepackage{amsmath}
\usepackage{amsfonts}
\usepackage{authblk}
\usepackage{xcolor}
\usepackage{url}

\begin{document}
%
\title{Self-supervised Depth Estimation to Regularise Semantic Segmentation in Knee Arthroscopy}
%
%
\author{Fengbei Liu\inst{1} \and
Yaqub Jonmohamadi \inst{2} \and
Gabriel Maicas\inst{1} \and
Ajay K. Pandey \inst{2} \and 
Gustavo Carneiro \inst{1}}
\authorrunning{Liu et al.}
%
\institute{Australian Institute for Machine Learning, School of Computer Science, University of Adelaide\\
\email{\{fengbei.liu,gabriel.maicas,gustavo.carneiro\}@adelaide.edu.au}\and
School of Electrical Engineering and Robotics, Science and Engineering Faculty, Queensland University of Technology \\
\email{\{y.jonmo,a2.pandey\}@qut.edu.au}\\}
\maketitle              
\begin{abstract}

Intra-operative automatic semantic segmentation of knee joint structures can assist surgeons during knee arthroscopy in terms of situational awareness.  However, due to poor imaging conditions (e.g., low texture, overexposure, etc.), automatic semantic segmentation is a challenging scenario, which justifies the scarce literature on this topic.  
In this paper, we propose a novel self-supervised monocular depth estimation to regularise the training of the semantic segmentation in knee arthroscopy. 
 To further regularise the depth estimation,  we propose the use of clean training images captured by the stereo arthroscope of routine objects (presenting none of the poor imaging conditions and with rich texture information) to pre-train the model. We fine-tune such model to produce both the semantic segmentation and self-supervised monocular depth using stereo arthroscopic images taken from inside the knee. 
Using a data set containing 3868 arthroscopic images captured during cadaveric knee arthroscopy with semantic segmentation annotations, 2000 stereo image pairs of cadaveric knee arthroscopy, and 2150 stereo image pairs of routine objects, we show that our semantic segmentation regularised by self-supervised depth estimation produces a more accurate segmentation than a state-of-the-art semantic segmentation approach modeled exclusively with semantic segmentation annotation.

\end{abstract}
%
%


\section{Introduction}

Knee arthroscopy is a minimally invasive surgery (MIS) conducted via small incisions that reduce surgical trauma and post-operation recovery time~\cite{siemieniuk2017arthroscopic}.  Despite these advantages, arthroscopy has some drawbacks, namely: limited access and loss of direct eye contact with the surgical scene, limited field of view (FoV) of the arthroscope, tissues too close to the camera (e.g., 10 mm away) being only partially visible in the camera FoV, diminished hand-eye coordination, and prolonged learning curves and training periods~\cite{smith2012advanced}.  In this scenario, surgeons can only confidently identify the femur due to its distinctive shape, while other structures, such as meniscus, tibia, and anterior cruciate ligament (ACL), remain challenging to be recognised. 
This limitation increases surgical operation time and may lead to unintentional tissue damage due to un-tracked camera movements. 
The automatic segmentation of these tissues has the potential to help surgeons by providing contextual awareness of the surgical scene, reducing surgery time, and decreasing the learning curve~\cite{price2015evidence}.


\begin{figure}[t!]
    \centering
    \includegraphics[width=0.6\linewidth]{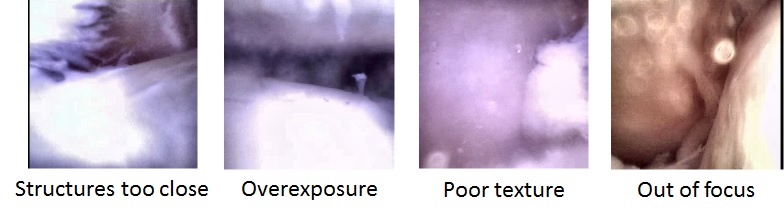}
    \caption{Challenging imaging conditions from knee arthroscopy.}
    \label{fig:image_challenges}
\end{figure}
Deep learning (DL) semantic segmentation has been intensively studied by the computer vision community~\cite{long2015fully,badrinarayanan2017segnet,lin2017refinenet,ronneberger2015u,chen2017deeplab}. For arthroscopy, we are aware of just one method that produces automatic semantic segmentation of knee structures~\cite{jonmohamadi2020automatic}. 
These semantic segmentation approaches tend to be prone to overfitting, depending on the data set available for the training process.
As a consequence, there is an increasing interest in the development of regularisation methods, such as the ones based on multi-task learning (MTL)~\cite{ruder2017overview}. For instance, fusing semantic segmentation and depth estimation has been shown to be an effective approach~\cite{eigen2015predicting}, but it requires the manual annotation for the training of the  segmentation and depth tasks. Considering that obtaining the depth ground truth for knee arthroscopy is challenging, self-supervised techniques such as~\cite{garg2016unsupervised,godard2017unsupervised} are highly favourable as they do not require ground truth depth. A similar approach has been successfully explored in robotic surgery~\cite{ye2017self}, but not for knee arthroscopy.  Moreover, self-supervised depth estimation techniques have been recently combined with semantic segmentation for training regularisation in non-medical imaging approaches~\cite{chen2019towards,ramirez2018geometry}. 
Nevertheless, these approaches rely on data sets that contain stereo images captured from street or indoor scenes, where visual objects are far from the camera, contain rich texture, and  images have few recording issues, such as overexposure and focus problems. 
On the other hand, knee arthroscopy images generally suffer from under or overexposure and focus problems, where visual objects are too close to the camera and contain poor texture, as shown in Fig.~\ref{fig:image_challenges}.

\begin{figure}[t!]
    \centering
    \includegraphics[width=.8\linewidth]{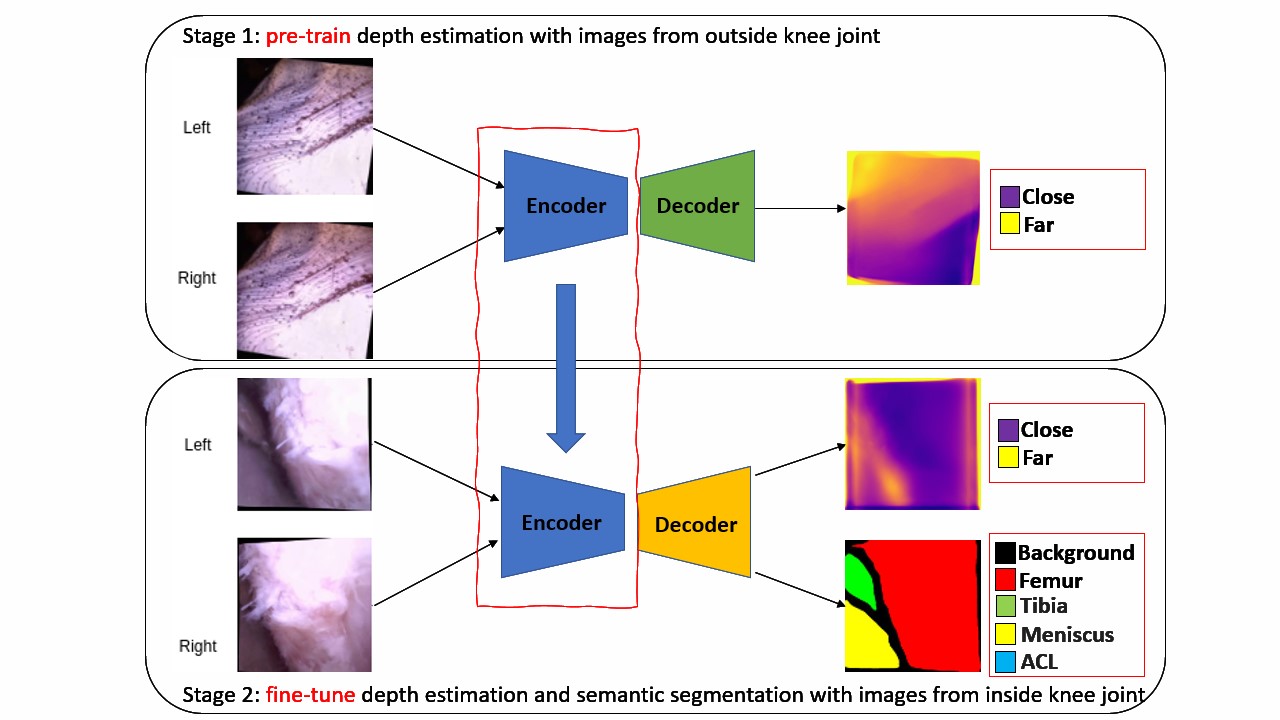}
    \caption{The proposed method is first pre-trained with the self-supervised depth estimation using stereo arthroscopic images of routine objects, where the images contain none of the issues of Fig.~\ref{fig:image_challenges}.  Stage two fine-tunes the model by training a fully supervised semantic segmentation  regularised by a self-supervised depth estimation. The output contains the segmentation mask and the depth estimation for the arthroscopic image. A detailed structure of encoder and decoder are shown in Fig.~\ref{fig:model}.}
    \label{fig:model_motivation}
\end{figure}

In this paper, we present an MTL approach for jointly estimating semantic segmentation and depth, where our aim is to use self-supervised depth estimation from stereo images to regularise the semantic segmentation training from knee arthroscopy.
Contrary to~\cite{ramirez2018geometry} that uses outdoor scenes, we tackle the segmentation of challenging arthroscopy images (Fig.~\ref{fig:image_challenges}). 
To this end, we pre-train our model on images of routine objects that do not show any of the issues displayed in Fig.~\ref{fig:image_challenges}.  
Then, we fine-tune our model with an MTL loss formed by the fully supervised semantic segmentation and the self-supervised depth estimation, as shown in Fig.~\ref{fig:model_motivation}. Using a data set containing 3868 arthroscopic images (with semantic segmentation annotations), 2000 stereo pairs captured during five cadaveric experiments and 2150 stereo image pairs of routne objects, we demonstrate that our method achieves higher accuracy in semantic segmentation (for the visual classes Femur, Meniscus, Tibia, and ACL) than state-of-the-art pure semantic segmentation methods.

\section{Proposed Method}

\subsection{Data Sets}

We use three data sets: 1) the pre-training depth estimation data set $\mathcal{D}^{pre} = \{ (\mathbf{I}^l, \mathbf{I}^r)_{k,n}\}_{k=1,n=1}^{|\mathcal{D}^{pre}|,N^{pre}_k}$, where $l$ and $r$ represent the left and right images of a stereo pair, $k$ indexes the out-of-the-knee scene, and $N^{pre}_k$ denotes the number of frames in the $k^{th}$ scene; 
2) the fine-tuning depth estimation and semantic segmentation data sets, respectively denoted by $\mathcal{D}^{dep} = \{ (\mathbf{I}^l, \mathbf{I}^r)_{k,n}\}_{k=1,n=1}^{|\mathcal{D}^{dep}|,N^{dep}_k}$ and $\mathcal{D}^{seg} = \{ (\mathbf{I}, \mathbf{y})_{k,n}\}_{k=1,n=1}^{|\mathcal{D}^{seg}|,N^{seg}_k}$, where $k$ indexes a human knee, and $N^{pre}_k$ and $N^{seg}_k$ denote the number of frames in the $k^{th}$ knee.
In these data sets, colour images are denoted by $\mathbf{I}:\Omega \rightarrow \mathbb{R^{3}}$,where $\Omega$ represents the image lattice, and the semantic annotation is represented by $\mathbf{y}:\Omega \rightarrow \mathcal{Y}$, with $\mathcal{Y}=\{\text{Background, Femur, Tibia, Meniscus, ACL}\}$. 

\subsection{Data Set Acquisition}

\begin{figure}[t!]
    \centering
    \includegraphics[width=.45\linewidth]{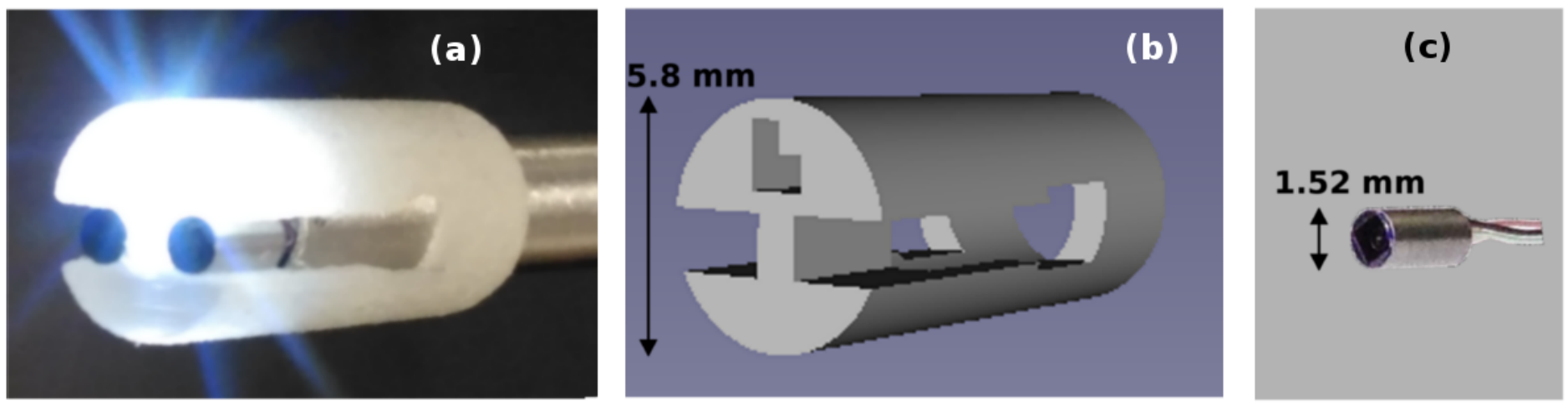}
    \caption{The custom built camera. The camera tip is shown in (a), the 3D design is displayed in (b), and the muC103A camera is in (c).}
    \label{fig:camera}
\end{figure}

\begin{table}[t!]
\caption{Percentage of training images per cadaver containing each of the structures~\cite{jonmohamadi2020automatic}.}
\label{tab:label_information}
\centering
\resizebox{0.6\columnwidth}{!}{%
\begin{tabular}{|c|c|c|c|c|c|}
\hline
\textbf{\begin{tabular}[c]{@{}c@{}}Structure\\ Cadaver knee\end{tabular}} & \textbf{Femur} & \textbf{ACL}  & \textbf{Tibia} & \textbf{Meniscus} & \textbf{Number of images} \\ \hline
\textbf{1}                                                                & \textbf{40\%}  & \textbf{0\%}  & \textbf{7\%}   & \textbf{0\%}      & \textbf{99}               \\ \hline
\textbf{2}                                                                & \textbf{32\%}  & \textbf{20\%} & \textbf{5\%}   & \textbf{9\%}      & \textbf{1043}             \\ \hline
\textbf{3}                                                                & \textbf{30\%}  & \textbf{14\%} & \textbf{8\%}   & \textbf{10\%}     & \textbf{1768}             \\ \hline
\textbf{4-left}                                                           & \textbf{47\%}  & \textbf{3\%}  & \textbf{4\%}   & \textbf{6\%}      & \textbf{459}              \\ \hline
\textbf{4-right}                                                          & \textbf{33\%}  & \textbf{8\%}  & \textbf{9\%}   & \textbf{12\%}     & \textbf{489}              \\ \hline
\textbf{Total}                                                            & \textbf{33\%}  & \textbf{13\%} & \textbf{7\%}   & \textbf{9\%}      & \textbf{3868}             \\ \hline
\end{tabular}
}
\end{table}
The arthroscopy images were acquired with a monocular Stryker endoscope
(4.0 mm diameter) and a custom built stereo arthroscope
using two muC103A cameras and a white LED for illumination (see Fig.~\ref{fig:camera}). 
The Stryker endoscope has resolution $1280 \times 720$ with  FoV of 30 degrees, and the custom built camera has resolution $384 \times 384$ and FoV of 87.5 degrees. 
Stryker images were cropped to have resolution $720 \times 720$ and then down-sampled to $384 \times 384$. 
Two clinicians performed the semantic segmentation annotations for classes femur, ACL, tibia and meniscus of 3868 images taken from four cadavers (where for one of the cadavers we used images from both knees) -- see annotation details in Tab.~\ref{tab:label_information}.

We also collected 2000 stereo pairs captured during these five cadaveric experiments.
The data set with images acquired of routine objects 
contains 2050 stereo images pairs 
used for pre-training the depth estimator and 100 stereo image pairs to validate the depth output performance (see an example of this type of image in Fig.~\ref{fig:model_motivation}). 
To fine tune the depth estimation method, we grab video frames from original arthroscopy stereo camera video by every two seconds. 

Note that there is no disparity ground truth available for any of the data sets above, so we cannot estimate the performance of the depth estimator.
\subsection{Model for Semantic Segmentation and Self-supervised Depth Estimation}

\begin{figure}[t!]
    \centering
    \includegraphics[width=.95\linewidth]{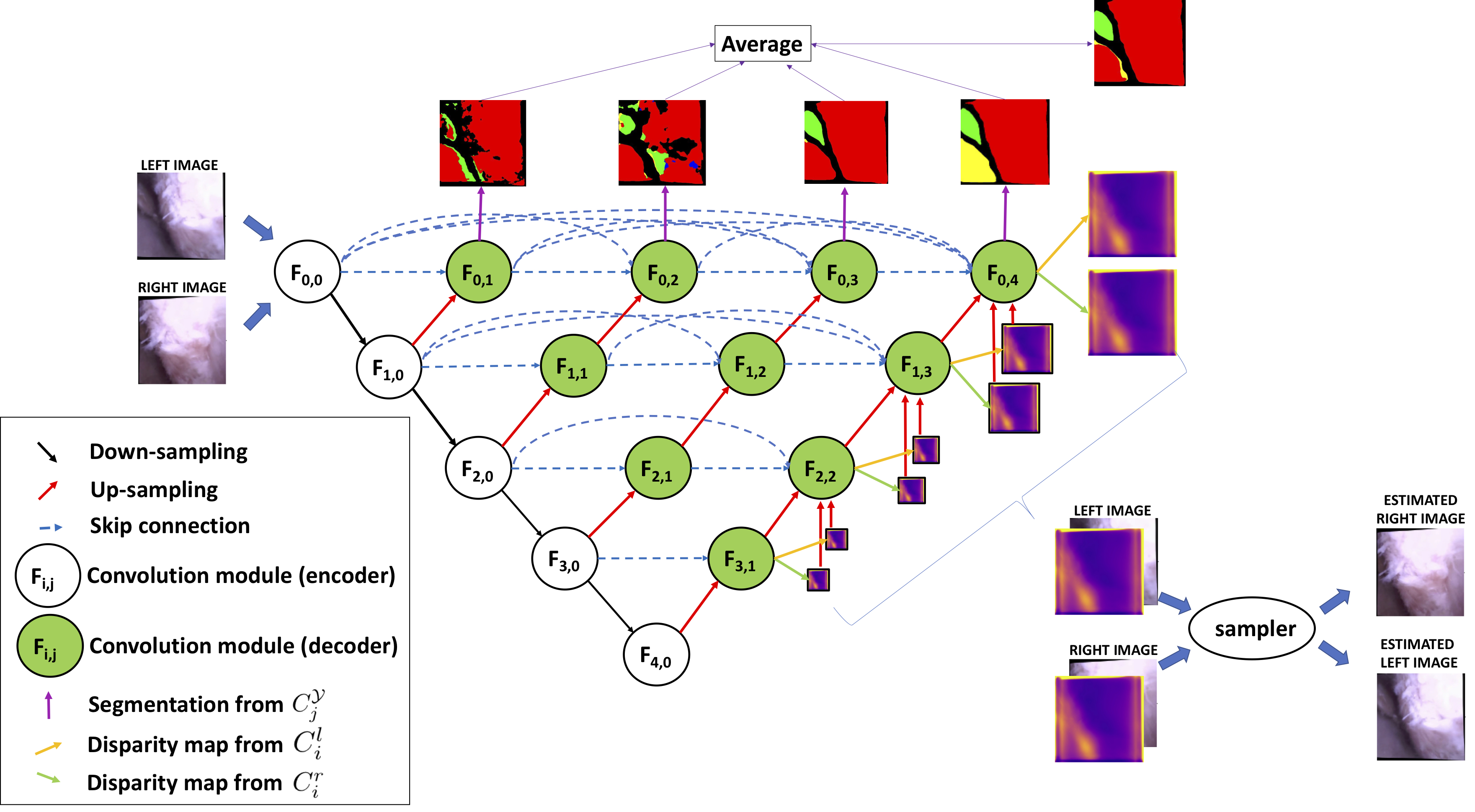}
    \caption{We extend Unet++~\cite{zhou2018unet++} to produce  multi-level semantic segmentation (on top), and multi-resolution
    disparity (inverse depth) estimations between the left and right images from the stereo pair (on the right hand side).  During training, all levels and resolutions of semantic segmentation and depth estimation are used, and for testing, we only output the result from the finest segmentation level and depth resolution.}
    \label{fig:model}
\end{figure}

The goal of our proposed network is to simultaneously estimate semantic segmentation and depth estimation from a single image. 
Motivated by~\cite{jonmohamadi2020automatic}, the model backbone is the U-net++~\cite{zhou2018unet++}, which predicts semantic segmentation and depth at four different levels, where the features are shared between these two tasks, as shown in Fig~\ref{fig:model_motivation}.


In the model depicted in Fig.~\ref{fig:model}, each module $F_{i,j}(\mathbf{x};\theta_{i,j})$ consists of blocks of convolutional layers (the input is represented by $\mathbf{x}$ and weights are represented by $\theta_{i,j}$), 
where the index $i$ denotes the down-sampling layer and $j$
represents the convolution layer of the dense block along the same skip connections (horizontally in the model). 
These modules are defined by
\begin{equation}
\mathbf{x}_{i,j} = 
\begin{cases} 
F_{i,j}(\mathbf{x}_{i-1,j};\theta_{i,j})& \quad \text{if }j=0 \\ 
F_{i,j}([[\mathbf{x}_{i,l}]_{l=0}^{j-1},U(\mathbf{x}_{i+1,j-1}),U(\mathbf{d}_i^l),U(\mathbf{d}_i^r)];\theta_{i,j})& \quad \text{if }j>0
\end{cases},
\label{eq:conv_module}
\end{equation}
where $U(.)$ denotes an up-sampling layer (using bilinear interpolation), $[.]$ represents a concatenation layer, and $\mathbf{d}_i^{\{l,r\}}$ is the disparity map that is defined only when $(i,j) \in \{(2,2),(1,3),(0,4) \}$ (otherwise it is empty), as described below in Eq.~\ref{eq:disparity_map}. The input image $\mathbf{I}$ enters the model at $F_{0,0}(\mathbf{I},\theta_{0,0})$.  
Each encoder convolution module (white nodes in Fig.~\ref{fig:model_motivation}) consists of a 
3 $\times$ 3 filter followed by max pooling, 
and each decoder convolution module (green nodes in Fig.~\ref{fig:model_motivation}) comprises bi-linear upsampling with scale factor 2, followed by two layers of 3 $\times $ 3 filters, batch normalization and ReLU. 
The semantic segmentation output consists of 
\begin{equation}
    \tilde{\mathbf{y}}_j = C_j^{\mathcal{Y}}(\mathbf{x}_{0,j};\theta^{\mathcal{Y}}_j),
    \label{eq:semantic_segmentation_map}
\end{equation}
where $j\in\{1,2,3,4\}$, and $C_j^{\mathcal{Y}}(.)$ is a convolutional layer parameterised by $\theta^{\mathcal{Y}}_j$ that outputs the estimation of the semantic segmentation $\tilde{\mathbf{y}}_j:\Omega \rightarrow \mathcal{Y}$ for the $j^{th}$ convolutional layer.  
In particular, $C_j^{\mathcal{Y}}(.)$ is formed by a 1 $\times$ 1 convolution filter followed by pixel-wise softmax activation. 
The left and right disparity maps are obtained from
\begin{equation}
    \mathbf{d}_i^{\{l,r\}} =
    C_{i}^{\{l,r\}}(\mathbf{x}_{i,j};\theta^{\{l,r\}}_{i})
    \label{eq:disparity_map}
\end{equation}
where $(i,j) \in \{(3,1),(2,2),(1,3),(0,4) \}$, and $C_{i}^{\{l,r\}}(.)$ is a convolutional layer parameterised by $\theta^{\{l,r\}}_{i}$ that outputs the estimation of the left and right disparity maps $\mathbf{d}_i^{\{l,r\}}:\Omega_i \rightarrow \mathbb R$ for the resolution at the $i^{th}$ down-sampling layer with $\Omega_i$ representing the image lattice at the same layer. The nodes $C_{i}^{\{l,r\}}(\mathbf{x}_{i,j};\theta^{\{l,r\}}_{i})$ consist of a 3 $\times$ 3 convolution filter with sigmoid activation to estimate the disparity result. 



The \textbf{training for the supervised semantic segmentation} for a particular image $\mathbf{I}^l$ with annotation $\mathbf{y}$ and the average
semantic segmentation results from the intermediate layers $ \bar{\mathbf{y}} = \sum_{j=1}^4 \tilde{\mathbf{y}}_j$ from ~\eqref{eq:semantic_segmentation_map} is based on the minimisation of the following loss function~\cite{milletari2016v}:
\begin{equation}
    \ell_{se}(\mathbf{y},\bar{\mathbf{y}}_j) = \alpha_{ce} \ell_{ce}(\mathbf{y},\bar{\mathbf{y}}) + (1 - \ell_{Dice}(\mathbf{y},\bar{\mathbf{y}})),
    \label{eq:semantic_semgmentation_loss}
\end{equation}
where $\ell_{ce}(\mathbf{y},\bar{\mathbf{y}})$ is the pixel-wise cross entropy loss computed between the annotation $\mathbf{y}$ and the average of the estimated semantic segmentation $\bar{\mathbf{y}}$,  $\ell_{Dice}(\mathbf{y},\bar{\mathbf{y}})$ denotes the Dice loss~\cite{milletari2016v}, with $\alpha_{ce}$ being set to $0.5$. 
The \textbf{inference for the supervised semantic segmentation} is based solely on the segmentation result from the last layer $\tilde{\mathbf{y}}_4$ from~\eqref{eq:semantic_segmentation_map}.


The \textbf{self-supervised depth estimation training}~\cite{godard2017unsupervised} uses rectified stereo pair images $\mathbf{I}^{\{l,r\}}$ to predict the disparity maps $\{ \mathbf{d}_i^{\{l,r\}} \}_{i=0}^3$ to match the left-to-right and right-to-left images.  The loss to be minimised is defined as \begin{equation}
\begin{split}
    \ell_d(\mathbf{I}^l,\mathbf{I}^r) = \sum_{i=0}^3 \Big  [ &
    \alpha_{ap} \Big ( \sum_{m\in\{l,r\}} \ell_{ap}^{m}(\mathbf{I}^l,\mathbf{I}^r,\mathbf{d}_i^m) \Big ) +
    \alpha_{lr} \Big ( \sum_{m\in\{l,r\}} \ell_{lr}^{m}(\mathbf{I}^l,\mathbf{I}^r,\mathbf{d}_i^m) \Big ) + \\
    & \alpha_{ds} \Big ( \sum_{m\in\{l,r\}} \ell_{ds}^{m}(\mathbf{I}^l,\mathbf{I}^r,\mathbf{d}_i^m) \Big )  \Big ],
\end{split}
\label{eq:disparity_loss}
\end{equation}
where
\begin{equation}
\ell_{ap}^{l}(\mathbf{I}^l,\mathbf{I}^r,\mathbf{d}_i^l) = \frac{1}{|\Omega_i|} \Big [ \sum_{\omega \in \Omega_i} \Big ( \gamma \Big (  \frac{1-SSIM(\mathbf{I}^l({\omega}),\tilde{\mathbf{I}}^l({\omega}))}{2} \Big ) + (1-\gamma)| \mathbf{I}^l({\omega}) - \tilde{\mathbf{I}}^l({\omega}) | \Big ) \Big ] ,
\end{equation}
where $\ell_{ap}^{r}(\mathbf{I}^l,\mathbf{I}^r,\mathbf{d}_i^r)$ is similarly defined, $SSIM(.)$ represents the structural similarity index~\cite{wang2004image}, $|\Omega_i|$ denotes the size of the image lattice at the $i^{th}$ resolution, $\tilde{\mathbf{I}}^l$ is the reconstructed left image using the right image re-sampled from the disparity map $\mathbf{d}_i^l$.  Also in~\eqref{eq:disparity_loss}, we have
\begin{equation}
    \ell_{lr}^l(\mathbf{I}^l,\mathbf{I}^r,\mathbf{d}_i^l) = \sum_{\omega \in \Omega_i}  \Big |  \mathbf{d}^l_i({\omega}) - \mathbf{d}^l_i({\omega + \mathbf{d}_i^r({\omega})}) \Big |,
\end{equation}
and similarly for $\ell_{lr}^r(\mathbf{I}^l,\mathbf{I}^r,\mathbf{d}_i^r)$ -- this loss minimises the $\ell_1$-norm between the left disparity map $\mathbf{d}_i^l$ and the transformed right-to-left disparity map. The last loss term in~\eqref{eq:disparity_loss} is defined by
\begin{equation}
    \ell_{ds}^{l}(\mathbf{I}^l,\mathbf{I}^r,\mathbf{d}_i^l) = \frac{1}{|\Omega_i|} \sum_{\omega \in \Omega_i} | \partial_x \mathbf{d}_i^l({\omega})  |\times e^{- \|\partial_x \mathbf{I}^l({\omega}) \|} + 
    | \partial_y \mathbf{d}_i^l({\omega})  |\times e^{- \|\partial_y \mathbf{I}^l({\omega}) \|},
\end{equation}
and similarly for $\ell_{ds}^{r}(\mathbf{I}^l,\mathbf{I}^r,\mathbf{d}_i^r)$ -- this loss penalises large disparity changes in smooth regions of the image, and when there are large image changes, there can be large transitions in the disparity maps. The \textbf{inference for the depth estimation} relies on the result for the finer scale $\mathbf{d}_0^{\{l,r\}}$.

\textbf{Model pre-training} is done with the data set $\mathcal{D}^{pre}$ by minimising the depth estimation loss~\eqref{eq:disparity_loss}, where we learn the model parameters $\{ \theta_{i,j} \}_{i,j \in \{0,1,2,3,4\}}$ in~\eqref{eq:conv_module} and disparity module parameters $\{\theta_i^{\{l,r\}}\}_{i \in \{0,1,2,3\}}$ in~\eqref{eq:disparity_map}.  After pre-training, we add the layers $\{ C_j^{\mathcal{Y}}\}_{j \in \{1,2,3,4\}}$ and perform an
\textbf{end-to-end training of all model parameters} with $\mathcal{D}$ by summing the losses in~\eqref{eq:semantic_semgmentation_loss} and~\eqref{eq:disparity_loss}.

\section{Experiments and Results}

\begin{figure}[!t]
    \centering
    \includegraphics[width=.9\linewidth]{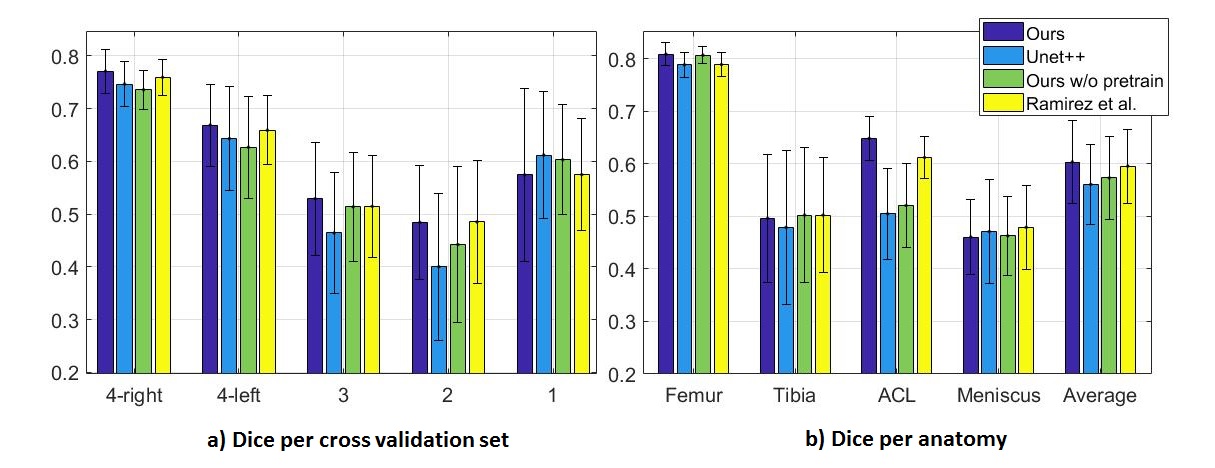}
    \caption{Dice results over each test set (left) and each anatomy (right), and the final average over all sets and anatomies (rightmost) for all methods tested in this paper.}
    \label{fig:results}
\end{figure}

We implement our model in Pytorch~\cite{paszke2017automatic}. 
The encoder for the model consists of the ResNet50~\cite{he2016deep}. 
\textbf{Pre-training} takes 200 epochs with batch size 32, where initial learning rate is $10^{-4}$ and halved at 80 and 120 epochs, and we use Adam~\cite{kingma2014adam} optimizer. Data augmentation includes random horizontal and vertical flipping, random gamma from [0.8,1.2], brightness [0.5,2.0], and colour shifts [0.8,1.2] by sampling from uniform distributions. 
For \textbf{fine-tuning} of segmentation and depth using arthroscopic images, we use the pre-trained encoder and re-initialise the decoder. The training takes 120 epochs with batch size 12. We use polynomial learning decay~\cite{zhao2017pyramid} with $\gamma = 0.9$ and weight decay $10^{-5}$. 
The data augmentation for segmentation includes horizontal and vertical flipping, random brightness contrast change and non-rigid transformation, including elastic transformation (the elastic transformation was particularly important to avoid over-fitting the training set) and depth data augmentation is the same as pre-training stage. For the inference time, the network takes 50ms to process a single test image and output the segmentation mask and depth.

\begin{figure}[!t]
    \centering
    \includegraphics[width=0.95\textwidth]{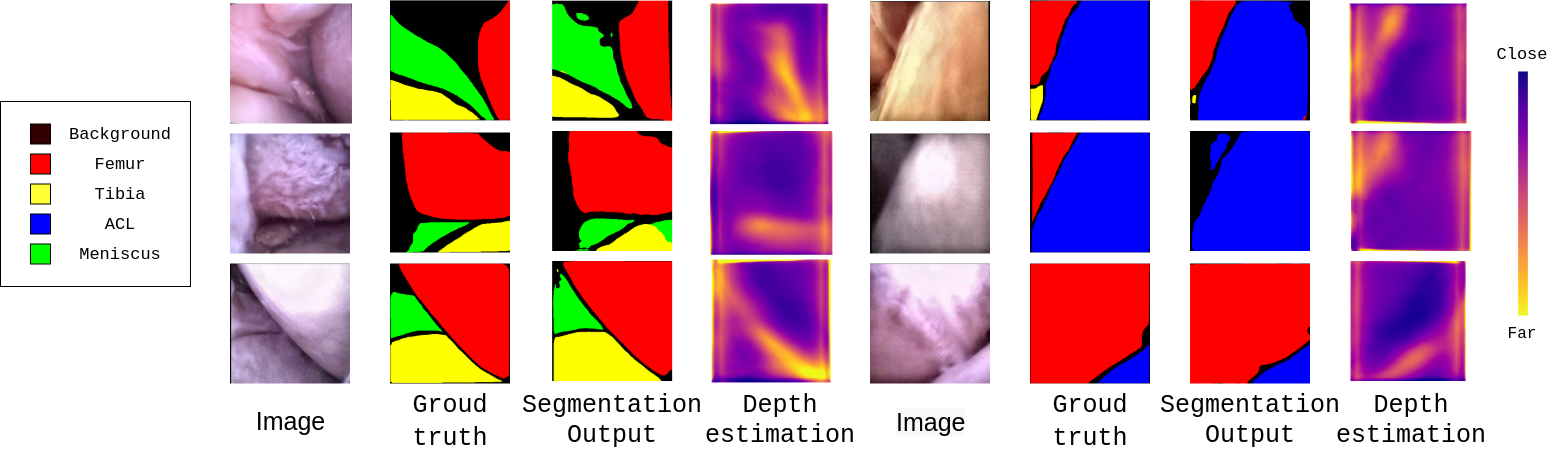}
    \caption{Examples of results, including original arthroscopy image, segmentation ground truth, proposed method segmentation prediction and unsupervised depth estimation.}
    \label{fig:test}
\end{figure}

We assess the performance of our method using the Dice coefficient computed on the testing set in a leave one out cross validation experiment (i.e., we train with 4 knees and test with the remaining one from Tab.~\ref{tab:label_information}).  In Fig.~\ref{fig:results} we show the mean and standard deviation of the Dice results over each test set and each anatomy, and the final average over all sets and anatomies.  We compare our newly proposed method (labelled as Ours) against the pure semantic segmentation model Unet++~\cite{zhou2018unet++,jonmohamadi2020automatic}, our method without the pre-training stage (labelled as Ours w/o pretrain), and the joint semantic segmentation and depth estimation method designed for computer vision applications by Ramirez et al.~\cite{ramirez2018geometry}.  The results indicate that our method (mean Dice of $0.603\pm0.159$) is significantly better than Unet++ (mean Dice of $0.560\pm0.152$), with a Wilcoxon signed rank test showing a p-value $\textless$ 0.05, indicating that the use of depth indeed improves the segmentation result from a pure segmentation method~\cite{zhou2018unet++,jonmohamadi2020automatic}.  In fact, our method produces significant gains in the segmentation of ACL (arguably the most challenging anatomy in the experiment).
Our method that uses pre-training is better than the one without pre-training (mean Dice of $0.573\pm0.157$), but not significantly so given that p-value is $\textgreater$ 0.05.  An interesting point is that even though our method without pre-training is better than the pure segmentation approach, it still cannot produce accurate segmentation for ACL.
Finally, compared to the method by Ramirez et al.~\cite{ramirez2018geometry} (mean Dice of $0.595\pm0.141$) ours is slightly better, indicating that both methods are competitive. 
Figure~\ref{fig:test} shows a few segmentation and depth estimation results. Note that we cannot validate depth estimation because we do not have ground truth available for it.

\vspace{-.1cm}
\section{Conclusion}
\vspace{-.1cm}

In this paper, we proposed a method to improve semantic segmentation using self-supervised depth estimation for arthroscopic images. Our network architecture is end-to-end trainable and does not require depth annotation. 
We also showed that the use of arthroscopic images of normal objects to pre-train the model can mitigate the challenging image conditions presented by this problem. By using geometry information, the model provides a slight improvement in terms of semantic segmentation accuracy. Future work will focus on improving the segmentation accuracy for the non-femural anatomies.

\section{Acknowledgements}
We acknowledge several technical discussions that influenced this paper with Ravi Garg and Adrian Johnston. This work was supported by the Australia India Strategic Research Fund (Project AISRF53820) and in part by the Australian Research
Council through under Grant DP180103232. The cadaver studies is covered by the Queensland University of Technology Ethics Approval
under project1400000856.
%
%
%
%
\bibliographystyle{splncs04}
\bibliography{sample}

\begin{thebibliography}{10}
\providecommand{\url}[1]{\texttt{#1}}
\providecommand{\urlprefix}{URL }
\providecommand{\doi}[1]{https://doi.org/#1}

\bibitem{badrinarayanan2017segnet}
Badrinarayanan, V., Kendall, A., Cipolla, R.: Segnet: A deep convolutional
  encoder-decoder architecture for image segmentation. IEEE transactions on
  pattern analysis and machine intelligence  \textbf{39}(12),  2481--2495
  (2017)

\bibitem{chen2017deeplab}
Chen, L.C., Papandreou, G., Kokkinos, I., Murphy, K., Yuille, A.L.: Deeplab:
  Semantic image segmentation with deep convolutional nets, atrous convolution,
  and fully connected crfs. IEEE transactions on pattern analysis and machine
  intelligence  \textbf{40}(4),  834--848 (2017)

\bibitem{chen2019towards}
Chen, P.Y., Liu, A.H., Liu, Y.C., Wang, Y.C.F.: Towards scene understanding:
  Unsupervised monocular depth estimation with semantic-aware representation.
  In: Proceedings of the IEEE Conference on Computer Vision and Pattern
  Recognition. pp. 2624--2632 (2019)

\bibitem{eigen2015predicting}
Eigen, D., Fergus, R.: Predicting depth, surface normals and semantic labels
  with a common multi-scale convolutional architecture. In: Proceedings of the
  IEEE international conference on computer vision. pp. 2650--2658 (2015)

\bibitem{garg2016unsupervised}
Garg, R., BG, V.K., Carneiro, G., Reid, I.: Unsupervised cnn for single view
  depth estimation: Geometry to the rescue. In: European Conference on Computer
  Vision. pp. 740--756. Springer (2016)

\bibitem{godard2017unsupervised}
Godard, C., Mac~Aodha, O., Brostow, G.J.: Unsupervised monocular depth
  estimation with left-right consistency. In: Proceedings of the IEEE
  Conference on Computer Vision and Pattern Recognition. pp. 270--279 (2017)

\bibitem{he2016deep}
He, K., Zhang, X., Ren, S., Sun, J.: Deep residual learning for image
  recognition. In: Proceedings of the IEEE conference on computer vision and
  pattern recognition. pp. 770--778 (2016)

\bibitem{jonmohamadi2020automatic}
Jonmohamadi, Y., Takeda, Y., Liu, F., Sasazawa, F., Maicas, G., Crawford, R.,
  Roberts, J., Pandey, A.K., Carneiro, G.: Automatic segmentation of multiple
  structures in knee arthroscopy using deep learning. IEEE Access  \textbf{8},
  51853--51861 (2020)

\bibitem{kingma2014adam}
Kingma, D.P., Ba, J.: Adam: A method for stochastic optimization. arXiv
  preprint arXiv:1412.6980  (2014)

\bibitem{lin2017refinenet}
Lin, G., Milan, A., Shen, C., Reid, I.: Refinenet: Multi-path refinement
  networks for high-resolution semantic segmentation. In: Proceedings of the
  IEEE conference on computer vision and pattern recognition. pp. 1925--1934
  (2017)

\bibitem{long2015fully}
Long, J., Shelhamer, E., Darrell, T.: Fully convolutional networks for semantic
  segmentation. In: Proceedings of the IEEE conference on computer vision and
  pattern recognition. pp. 3431--3440 (2015)

\bibitem{milletari2016v}
Milletari, F., Navab, N., Ahmadi, S.A.: V-net: Fully convolutional neural
  networks for volumetric medical image segmentation. In: 2016 Fourth
  International Conference on 3D Vision (3DV). pp. 565--571. IEEE (2016)

\bibitem{paszke2017automatic}
Paszke, A., Gross, S., Chintala, S., Chanan, G., Yang, E., DeVito, Z., Lin, Z.,
  Desmaison, A., Antiga, L., Lerer, A.: Automatic differentiation in pytorch
  (2017)

\bibitem{price2015evidence}
Price, A., Erturan, G., Akhtar, K., Judge, A., Alvand, A., Rees, J.:
  Evidence-based surgical training in orthopaedics: how many arthroscopies of
  the knee are needed to achieve consultant level performance? The bone \&
  joint journal  \textbf{97}(10),  1309--1315 (2015)

\bibitem{ramirez2018geometry}
Ramirez, P.Z., Poggi, M., Tosi, F., Mattoccia, S., Di~Stefano, L.: Geometry
  meets semantics for semi-supervised monocular depth estimation. In: Asian
  Conference on Computer Vision. pp. 298--313. Springer (2018)

\bibitem{ronneberger2015u}
Ronneberger, O., Fischer, P., Brox, T.: U-net: Convolutional networks for
  biomedical image segmentation. In: International Conference on Medical image
  computing and computer-assisted intervention. pp. 234--241. Springer (2015)

\bibitem{ruder2017overview}
Ruder, S.: An overview of multi-task learning in deep neural networks. arXiv
  preprint arXiv:1706.05098  (2017)

\bibitem{siemieniuk2017arthroscopic}
Siemieniuk, R.A., Harris, I.A., Agoritsas, T., Poolman, R.W.,
  Brignardello-Petersen, R., Van~de Velde, S., Buchbinder, R., Englund, M.,
  Lytvyn, L., Quinlan, C., et~al.: Arthroscopic surgery for degenerative knee
  arthritis and meniscal tears: a clinical practice guideline. Bmj
  \textbf{357},  j1982 (2017)

\bibitem{smith2012advanced}
Smith, R., Day, A., Rockall, T., Ballard, K., Bailey, M., Jourdan, I.: Advanced
  stereoscopic projection technology significantly improves novice performance
  of minimally invasive surgical skills. Surgical endoscopy  \textbf{26}(6),
  1522--1527 (2012)

\bibitem{wang2004image}
Wang, Z., Bovik, A.C., Sheikh, H.R., Simoncelli, E.P.: Image quality
  assessment: from error visibility to structural similarity. IEEE transactions
  on image processing  \textbf{13}(4),  600--612 (2004)

\bibitem{ye2017self}
Ye, M., Johns, E., Handa, A., Zhang, L., Pratt, P., Yang, G.Z.: Self-supervised
  siamese learning on stereo image pairs for depth estimation in robotic
  surgery. arXiv preprint arXiv:1705.08260  (2017)

\bibitem{zhao2017pyramid}
Zhao, H., Shi, J., Qi, X., Wang, X., Jia, J.: Pyramid scene parsing network.
  In: Proceedings of the IEEE conference on computer vision and pattern
  recognition. pp. 2881--2890 (2017)

\bibitem{zhou2018unet++}
Zhou, Z., Siddiquee, M.M.R., Tajbakhsh, N., Liang, J.: Unet++: A nested u-net
  architecture for medical image segmentation. In: Deep Learning in Medical
  Image Analysis and Multimodal Learning for Clinical Decision Support, pp.
  3--11. Springer (2018)

\end{thebibliography}

\end{document}